\documentclass[aps, prl, amsmath, showpacs, preprintnumbers, groupedaddress, twocolumn, amssymb, a4paper]{revtex4-1}
\usepackage{graphicx}
\usepackage{epstopdf}
\usepackage{mathptmx, textcomp}
\usepackage[latin1]{inputenc}
\usepackage[T1]{fontenc}

\begin{document}
\title{Probing the axis alignment of an ultracold spin-polarized $\textrm{Rb}_2$ molecule}
\author{Markus Dei{\ss}}
\author{Bj\"orn Drews}
\author{Benjamin Deissler}
\author{Johannes Hecker Denschlag}
\affiliation{Institut f\"ur Quantenmaterie and Center for Integrated Quantum Science and Technology IQ$^{ST}$, Universit\"at Ulm, 89069 Ulm, Germany}

\date{\today}

\begin{abstract}
We present a novel method for probing the alignment of the molecular axis of an ultracold, nonpolar dimer. These results are obtained using diatomic $^{87}$Rb$_2$ molecules in the vibrational ground state of the lowest triplet potential $a^3\Sigma_u^+$ trapped in a 3D optical lattice. We measure the molecular polarizabilities, which are directly linked to the alignment, along each of the $x$, $y$, and $z$ directions of the lab coordinate system. By preparing the molecules in various, precisely defined rotational quantum states we can control the degree of alignment of the molecular axis with high precision over a large range. Furthermore, we derive the dynamical polarizabilities for a laser wavelength of $1064.5\:\textrm{nm}$ parallel and orthogonal to the molecular axis of the dimer, $\alpha_\parallel=(8.9 \pm 0.9)\times10^3\:\textrm{a.u.}$ and $\alpha_\perp=(0.9 \pm 0.4)\times10^3\:\textrm{a.u.}$, respectively. Our findings highlight that the depth of an optical lattice strongly depends on the rotational state of the molecule which has to be considered in collision experiments. The present work paves the way for reaction studies between aligned molecules in the ultracold temperature regime. 
\end{abstract}

\pacs{33.15.Kr, 33.15.Bh, 37.10.Pq, 67.85.-d}

\maketitle

In molecular physics and chemistry, control over alignment or orientation of the molecular axis in the laboratory frame is often essential for understanding reaction processes and molecular structures (see, e.g., Refs. \cite{Stap03, Zare98, Krausz09, Ning12, Gijsbertsen07}). Currently, such experiments are typically carried out with molecular beams in a pulsed fashion, where the alignment or orientation is achieved by state selection with hexapole fields, optical preparation techniques, or exposure to strong ac or dc electric or magnetic fields (see, e.g., Refs. \cite{Baugh1994, Mukherjee11, Friedrich92, Lemeshko13, Stap03}). In general, the alignment of the molecular axis is measured via photodissociation, where the angular dependence of the fragments with respect to the laser polarization is measured.

A different approach entails working with optically trapped, ultracold molecular ensembles \cite{Krems08}. Such systems allow for extraordinary control over the internal and external degrees of freedom including the tailoring of the trapping potential and the preparation of molecules in precisely defined quantum states. Here, novel experimental regimes can be reached featuring ultralow-energy collisions and possible interaction times up to many seconds. Further prospects are reaction studies in reduced dimensions and selective investigations of few-body collisions by controlling the number of aligned particles per trapping site. Hence, ultracold molecules will strongly complement the research with molecular beams. 

In terms of  orientation, the first experiments with ultracold molecules were performed in 2011 where polar KRb molecules were  exposed to a dc electric field  \cite{Miranda11}. Afterwards, their anisotropic polarizability was investigated in a 1D optical lattice of which the polarization was rotated \cite{Neyenhuis2012}.

In this Letter, we demonstrate a novel method to determine the alignment of the molecular axis. This method can be readily implemented in typical ultracold-molecule setups. It relies on the fact that the axis alignment is directly reflected in an anisotropy of the molecular polarizability. The molecules are trapped in a cubic 3D optical lattice with orthogonal polarizations. For each lattice beam, we measure the potential depth for a given light intensity from which we infer the dynamical polarizabilities in the three directions of space and therefore the alignment of the molecular axis. As an application, we briefly show how this technique can be used to spectroscopically investigate unknown molecular states.

Our experiments are performed with ultracold $\textrm{Rb}_2$ molecules in the vibrational ground state of the $a^3\Sigma_u^+$ potential. The molecular ensemble is held in a 3D optical lattice at $1064.5\:\textrm{nm}$ with trapping times of several seconds. There is no more than a single molecule per lattice site. We are able to prepare a variety of precisely defined molecular energy eigenstates, where the rotational, Zeeman, and hyperfine structure is fully resolved. Consequently, hyperfine depolarization \cite{Rutkowski04, Bartlett10} plays no role. The $\textrm{Rb}_2$ molecules are $100\%$ spin polarized. The spin polarization directly determines the molecular axis alignment, which is the quantity that we measure in our experiment. Although no forced alignment via electric or magnetic fields is employed, sizeable degrees of alignment of the molecular axis are readily achieved. Furthermore, the alignment persists as long as a quantization axis is defined, in our case by an external magnetic field.

\begin{figure}
\includegraphics[width=8.0cm] {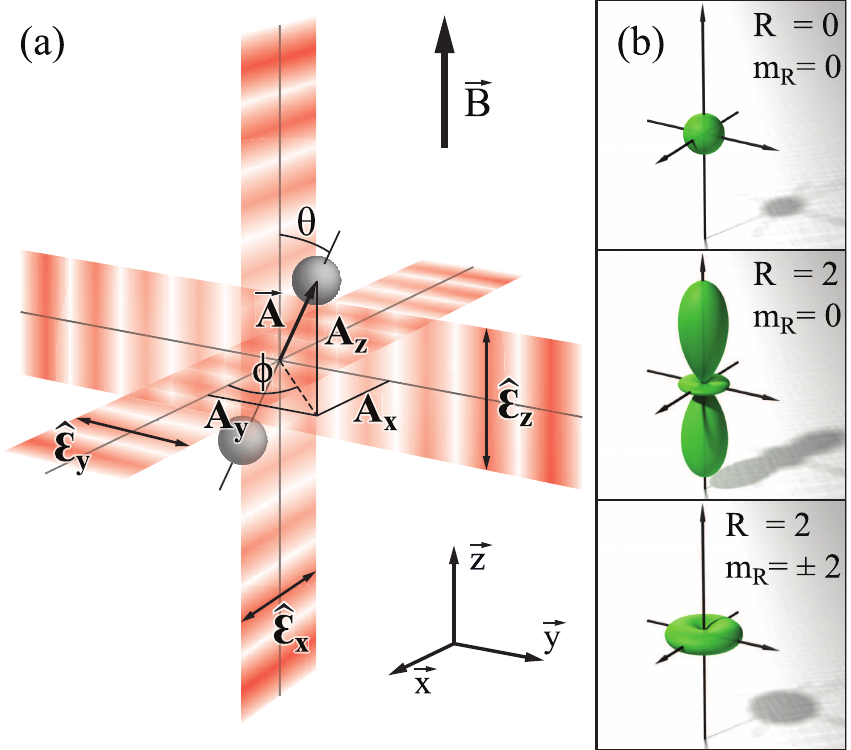}
\caption{(color online). (a) Schematic of the experiment. Three laser beams with linear polarizations $\hat{\epsilon}_i$ orthogonal to each other form a 3D optical lattice. The axis of a diatomic molecule is given by $\vec{A}$. The magnetic field $\vec{B}$ points in the $z$ direction and represents the quantization axis. (b) Polar plots of $|Y_{R,m_R}(\theta, \phi)|^2$ for states $|R,m_R\rangle = |0,0\rangle, |2,0\rangle, |2,\pm2\rangle$.
}
 \label{fig1}
\end{figure}

When a nonpolar molecule is exposed to a linearly polarized, oscillating electric field $\vec{E}(t)=\hat{\epsilon}E_0\cos(\omega t)$ with amplitude $E_0$ and unit polarization vector $\hat{\epsilon}$, a dipole potential $U = - \hat{\epsilon} \cdot (\tensor{\alpha} \hat{\epsilon} )  E_0^2/4 $ is induced. In a Cartesian coordinate system of which one axis is pointing along the molecular axis the polarizability tensor $\tensor{\alpha}$ of a dimer is diagonal and its components have two values $\alpha_\parallel$ and $\alpha_\perp$ for the directions parallel and perpendicular to the molecular axis. $U$ can then be written as $U = -(\alpha_\parallel E_\parallel^2 + \alpha_\perp E_\perp^2)/4$ where $E_\parallel$ and $E_\perp$ are the corresponding components of the electric field amplitude. We describe the orientation of the molecular axis by a unit vector $\vec{A}=(A_x,A_y,A_z)=(\textrm{sin}\theta \textrm{cos}\phi,\textrm{sin}\theta \textrm{sin}\phi,\textrm{cos}\theta)$, see Fig.$\:$\ref{fig1}(a). Using $E_\parallel^2 = (\hat{\epsilon} \cdot\vec{A})^2 E_0^2$ and $E_\perp^2 =  E_0^2 - E_\parallel^2$ the potential becomes $U = -\left[\alpha_\parallel (\hat{\epsilon}\cdot \vec{A})^2 + \alpha_\perp [1 - (\hat{\epsilon}\cdot \vec{A})^2] \right] E_0^2 /4 $. In a quantum mechanical treatment $(\hat{\epsilon} \cdot \vec{A})^2$ is replaced by its expectation value $\langle (\hat{\epsilon} \cdot \vec{A})^2 \rangle$. Using electric fields with amplitudes $E_{0,i}$ and polarizations $\hat{\epsilon}_i$ that point in each of the directions $(i=x,y,z)$ of the lab coordinate system we measure the molecular polarizabilities
 \begin{equation}
 \alpha^{(i)} = 4|U_i| /E_{0,i}^2 ,
 \label{Eq1}
 \end{equation}
  where
\begin{equation}
\alpha^{(i)}=\langle A_i^2 \rangle\alpha_\parallel+(1-\langle A_i^2 \rangle)\alpha_\perp.
\end{equation}
The quantity $\langle A_i^2 \rangle$ defines the degree of alignment of the molecular axis with respect to the direction $\hat{\epsilon}_i$. For nonaligned molecules, all $\langle A_i^2 \rangle$ are equal to $1/3$. Clearly, by measuring the polarizability $\alpha^{(i)}$ we can directly determine $\langle A_i^2 \rangle$ once $\alpha_\parallel$ and $\alpha_\perp$ are known.

For the $a ^3\Sigma_u^+$ state of Rb$_2$ the dynamics of the molecular axis are well described by the wave function of a quantum rotor,  of which the Hamiltonian is essentially given by $\vec{R}^2$ with $\vec{R}$ being the operator for nuclear rotation (see Ref. \cite{coupling}). Especially for magnetic fields larger than $100\:\textrm{G}$, we can in general safely ignore coupling of $\vec{R}$  to any other spins or angular momenta. Therefore $R$ and its projection $m_R$ onto the quantization axis in the $z$ direction are the only relevant quantum numbers to describe the angular distribution of the molecular axis, turning the Rb$_2$ molecule into a simple and fundamental system to study alignment. Consequently, the eigenstates of the axial motion are the spherical harmonics $Y_{R,m_R}(\theta,\phi) \equiv |R,m_R\rangle$. The degree of alignment of the molecular axis with respect to the $x$, $y$, and $z$ directions can be calculated as
\begin{equation}
\langle A_i^2 \rangle=\int|Y_{R,m_R}|^2A_i^2  \ \sin(\theta)d\theta d\phi.
\end{equation}
Figure \ref{fig1}(b) shows polar plots of $|Y_{R,m_R}|^2$ corresponding to relevant rotational states. For $|R = 0, m_R =  0\rangle $ the axis direction is isotropic in space, indicating a nonaligned molecule. In contrast, the axis direction is anisotropic for $|2,0\rangle $ and $|2,\pm2\rangle $, which is directly reflected in degrees of alignment different from $1/3$.

Our experimental setup and the molecule preparation has been described in detail in Ref. \cite{Lang2008}. In brief, an ultracold thermal ensemble of spin-polarized $^{87}\textrm{Rb}$ atoms ($f_\textrm{a}=1$, $m_{f_a}=1$) is loaded into a 3D optical lattice and converted into Feshbach molecules at a magnetic field of $B = 1007.4\:\textrm{G}$. Each Feshbach molecule is nonrotating and has magnetic quantum number $m_F =2$ of total angular momentum $\vec{F}$. The total nuclear spin is a superposition of components $I =  1, 2, 3$. Using an optical two-photon process [stimulated Raman adiabatic passage (STIRAP)] at $B = 1000\:\textrm{G}$, we transfer the molecules to the vibrational ground state of the $a ^3\Sigma_u^+$ potential, ending up with a pure ensemble of $1.5\times 10^4$  molecules at a temperature of about $1\:\mu\textrm{K}$. As we use $\pi$-polarized light, $m_F = 2$ does not change. The intermediate STIRAP level is located in the $c^3\Sigma_g^+$ potential, has quantum number $I =3$, and is a mixture of different $R$. Concerning the final level we choose to populate either one of the well-defined states $|R,m_R\rangle =|0,0\rangle$ or $|2,0\rangle$ which are separated by about $1.9\:\textrm{GHz}$ (see Ref. \cite{Strauss2010}) by setting the relative detuning of the STIRAP lasers.  Both levels have quantum numbers $I=3$ and $f=2$ ($\vec{f} = \vec{S} + \vec{I}$), where $\vec{S}$ denotes the total electronic spin. Compared to molecular beam setups, the STIRAP pulse in ultracold atoms or molecules experiments is orders of magnitude longer (typically tens of microseconds). It therefore can usually resolve any molecular substructure and unambiguously populate any quantum state as long as the selection rules allow for it.

The molecules reside within the lowest Bloch band of the optical lattice, which consists of a superposition of three linearly polarized standing light waves in the $x$, $y$, and $z$ directions with polarizations orthogonal to each other, see Fig.\:\ref{fig1}(a). Each lattice beam has a wavelength of $\lambda=1064.5\:\textrm{nm}$, a linewidth of a few kilohertz, and relative intensity fluctuations of less than $10^{-3}$. In order to avoid interference effects, the frequencies of the standing waves are offset by about $100\:\textrm{MHz}$ relative to each other. At the location of the atomic sample, the waists ($1/e^2$ radii) of the lattice beams are about $130\:\mu$m and the maximum available power per beam is about $3.5\:\textrm{W}$.

\begin{figure}
\includegraphics[width=8.0cm] {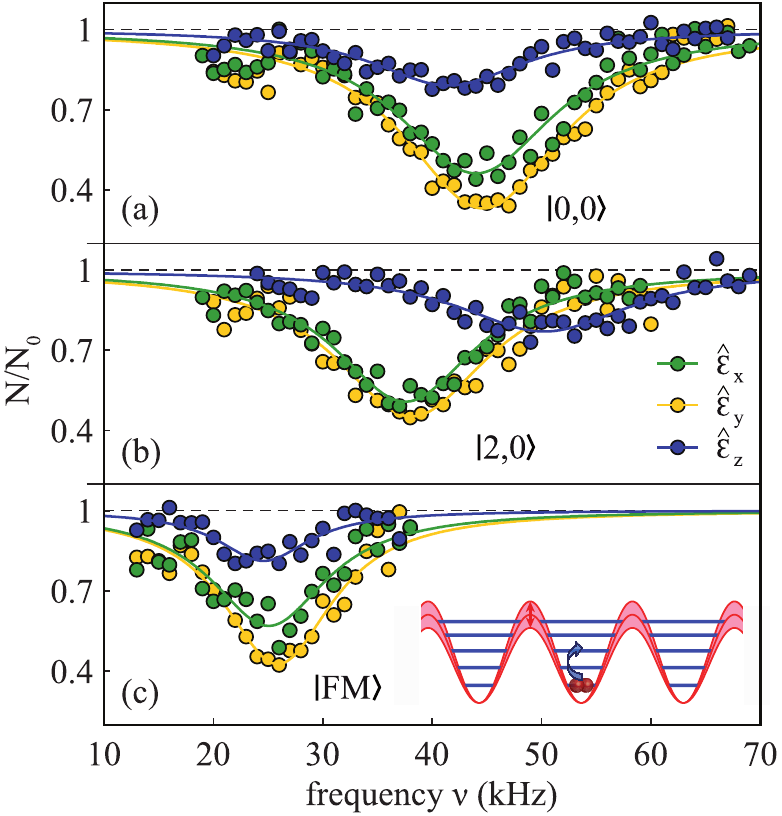}
\caption{(color online). Modulation spectroscopy. The general scheme is sketched in the inset in (c) illustrating the resonant modulation of the lattice depth in one direction. The data show resonances for molecules in states $|0,0\rangle$ (a) and $|2,0\rangle$ (b), as well as for Feshbach molecules $|\textrm{FM}\rangle$ (c) at $B=1000\:\textrm{G}$. We measured the fraction of remaining molecules $N/ N_0$ as a function of the modulation frequency $\nu$ for the three different lattice directions. Here, $\hat{\epsilon}_x$, $\hat{\epsilon}_y$, and $\hat{\epsilon}_z$ indicate the electric field of which the amplitude has been modulated [cf. Fig.\:\ref{fig1}(a)]. Each data point is the average of between five and 25 repetitions of the experiment (for a given molecular state and direction $i$ the number of repetitions is constant). The statistical error of each data point is typically $\pm(0.05-0.15)$. Solid lines correspond to Lorentzian fits.
}
 \label{fig2}
\end{figure}

We now independently determine the three molecular polarizabilities, $\alpha^{(i)} (i = x, y, z)$, which, according to Eq.\:(2), are directly linked to the degrees of molecular alignment $\langle A_{i}^2 \rangle$. Using Eq.\:(1), we need a measurement of both the lattice depth $U_i$ and the field amplitude $E_{0,i}$. In order to measure $U_i$ we consider a lattice beam with polarization $\hat{\epsilon}_i$ (see Fig.\:\ref{fig1}). We perform modulation spectroscopy  \cite{Denschlag02, Danzl2010} in which the intensity of this lattice beam is sinusoidally modulated by a few percent for a time of 10 modulation periods. A resonant modulation frequency drives transitions from the lowest Bloch band to the second excited band [see inset in Fig.\:\ref{fig2}(c)] giving rise to losses. At the end of each experimental cycle, the fraction $N/ N_0$ of molecules remaining is measured.  For this purpose, the atom signal is detected via absorption imaging after reversing the STIRAP and dissociating the molecules using a magnetic field sweep across the Feshbach resonance. By comparing the resonant transition frequency to a band-structure calculation of the sinusoidal lattice, the lattice depth $U$ is determined. 

Figure \ref{fig2} shows excitation spectra after modulation for vibrational ground state molecules in states $|0,0\rangle$ and $|2,0\rangle$ as well as for weakly bound Feshbach molecules at a magnetic field of $B=1000\:\textrm{G}$. For each measurement, we determine the center of the excitation resonance using a Lorentzian fit. For technical reasons the modulation strength varied between the three lattice directions ($6\%$, $5\%$, and $2\%$ peak-to-peak intensity modulation for $\hat{\epsilon}_x$, $\hat{\epsilon}_y$, and $\hat{\epsilon}_z$, respectively), leading to different depths of the resonances. We have checked that this variation does not affect the resonance positions. As the light field amplitudes $E_{0,i}$ for the three lattice directions are similar, we observe that for nonrotating molecules in $|0,0\rangle$ [Fig.\:\ref{fig2}(a)], the resonant excitation frequencies in the three different modulation directions are also very similar. For the case of $|2,0\rangle$ [Fig.\:\ref{fig2}(b)], however, the lattice depth for $\hat{\epsilon}_z$ polarization is much higher than for $\hat{\epsilon}_x$ and $\hat{\epsilon}_y$, indicating an alignment of molecules along the $z$ direction.

In order to precisely determine the electric field amplitudes $E_{0,i}$ we perform lattice modulation measurements with weakly bound Feshbach molecules for the same experimental parameters as for deeply bound molecules [Fig.\:\ref{fig2}(c)]. In this way, the polarizability can be determined independently of the exact beam parameters. We  use $\alpha_\mathrm{Rb+Rb}^{(i)} = 4|U_i|/E_{0,i}^2$ and the fact that the polarizability of Feshbach molecules, $\alpha_\textrm{Rb+Rb}$, is isotropic and known to be twice the atomic polarizability $\alpha_\textrm{Rb}= (693.5 \pm 0.9)\:\textrm{a.u.}$ \cite{Safronova2004}. Here, $1\:\mathrm{a.u.} = 4 \pi \epsilon_0 a_0^3 = 1.649 \times 10^{-41}\:\textrm{Jm}^2\textrm{V}^{-2}$, where $a_0$ denotes the Bohr radius and $\epsilon_0$ is the vacuum permittivity. As can be seen from the obtained excitation resonances [Fig.\:\ref{fig2}(c)], the absolute lattice depth and thus the electric field amplitude slightly varies in the three directions. Of course, according to Eq.\:(1), this variation drops out when determining the polarizabilities of the deeply bound molecules.

\begin{figure}
\includegraphics[width=8.0cm] {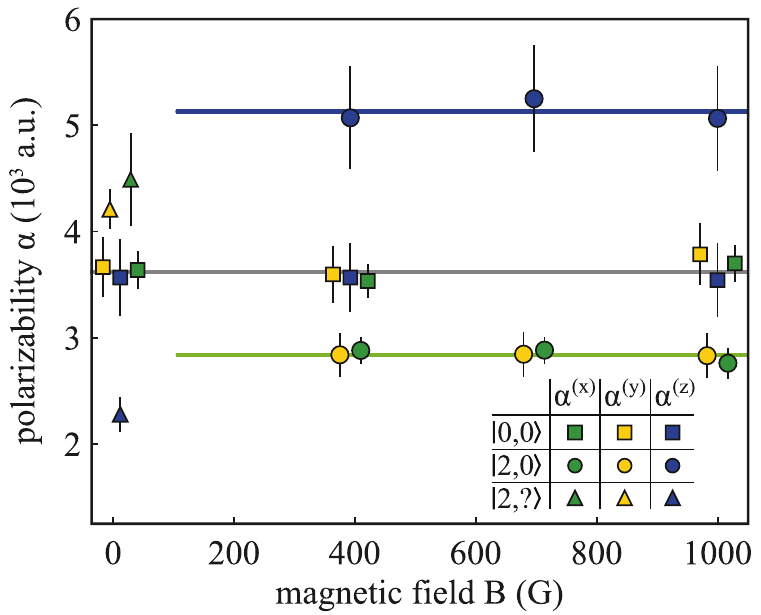}
\caption{(color online). Polarizabilities $\alpha^{(i)} (i = x, y, z)$ of the Rb$_2$ triplet molecules for the rotational states $|R= 0, m_R = 0\rangle$ (squares), $|2,0\rangle$ (circles), and an unknown state $|2,?\rangle$ (triangles) for different magnetic fields $B = 1000\:\textrm{G}$, $700\:\textrm{G}$, $400\:\textrm{G}$, and about $10\:\textrm{G}$. For better visibility overlapping data points are shifted slightly with respect to each other in the horizontal direction. The error bars are given by the uncertainty of the Lorentzian fits in the amplitude modulation spectra (cf. Fig.\:\ref{fig2}). Horizontal lines indicate the mean values of the measured polarizabilities for the states $|0,0\rangle$ and $|2,0\rangle$.
}
 \label{fig3}
\end{figure}

Figure \ref{fig3} shows measured polarizabilities $\alpha^{(i)}$ of molecules initially prepared in state $|0,0\rangle$ or $|2,0\rangle$ at $B = 1000\:\textrm{G}$. After production we adiabatically lower the magnetic field from $B = 1000\:\textrm{G}$ to $700\:\textrm{G}$, $400\:\textrm{G}$, and about $10\:\textrm{G}$. At each $B$ field we measure $\alpha^{(i)}$ in all three directions. As already seen in Fig.\:\ref{fig2}(a) the nonrotating state $|0,0\rangle$ exhibits an isotropic polarizability $\alpha^{(i)}$. This isotropy is reflected in the spherical symmetry of the rotational wave function for the molecular axis, $Y_{0,0}$, which is simply a constant. Correspondingly, the calculated expectation value $\langle A_i^2 \rangle$ is $1/3$ for all directions. Thus, Eq.\:(2) simplifies to the useful relation $3 \alpha^{(i)} = \alpha_\parallel + 2 \alpha_\perp$.

Next, we study $\alpha^{(i)}$ for state $|2,0\rangle$ down to $B = 400\:\textrm{G}$. As already observed in Fig.\:\ref{fig2}(b) there is a clear anisotropy of $\alpha^{(i)}$. The polarizabilities in the $x$ and $y$ directions are identical, but differ from the one in the $z$ direction. We check for the consistency of the measurements. From Eq.\:(2) and $\langle A_x^2 \rangle + \langle A_y^2 \rangle + \langle A_z^2 \rangle =1$ follows $\sum_{i} \alpha^{(i)}= \alpha_\parallel + 2 \alpha_\perp $, which should be equal to $3\alpha^{(i)}$ of state  $|0,0\rangle$. Our experimental data fulfill this relation to within $1\%$. The fact that the polarizability $\alpha^{(i)}$ is independent of $B$ both for $|0,0\rangle$ and $|2,0\rangle$ highlights that the alignment of the molecular axis is not forced by the magnetic field. It merely sets the direction of the quantization axis, stabilizing the spin polarization of the molecules.

We can use the measurements of $\alpha^{(i)}$ for state $|2,0\rangle = Y_{2,0}(\theta, \phi)$ to determine $\alpha_\parallel$ and $\alpha_\perp$ from Eq.\:(2). Using Eq.\:(3) we calculate $\langle A_x^2 \rangle = \langle A_y^2 \rangle = 0.2381$ and $\langle A_z^2 \rangle = 0.5238$ [cf. Fig.\:\ref{fig1}(b)]. We obtain a set of two independent equations (the equations for the $x$ and $y$ directions are nominally identical), which can be uniquely solved, resulting in $\alpha_\parallel=(8.9 \pm 0.9)\times10^3\:\textrm{a.u.}$ and $\alpha_\perp= (0.9 \pm 0.4)\times10^3\:\textrm{a.u.}$ These values are in good agreement with \textit{ab initio} calculations, which give $\alpha_\parallel=(7.5 \pm 1.2)\times10^3\:\textrm{a.u.}$ and $\alpha_\perp=(1.0 \pm 0.1)\times10^3\:\textrm{a.u.}$ \cite{TomzaPC,Tomza2013}. The large difference of  $\alpha_\parallel$ and $\alpha_\perp$ can be explained as mainly arising from the different lattice laser detuning with respect to the relevant electronic transitions \cite{supplemental}.

The novel method to probe the alignment of the molecular axis can be used to gain information about unknown quantum states and therefore has possible future applications in spectroscopy. As an example, we now look at the data points (triangles) at $10\:\textrm{G}$. Although these data are also obtained by first preparing state $|2,0\rangle$ at $1000\:\textrm{G}$ and subsequently ramping down the $B$ field, they look quite different from the ones at 400, 700, or $1000\:\textrm{G}$ discussed previously. All polarizabilities change considerably and $\alpha^{(z)}$ is now smaller than $\alpha^{(x)}$ and $\alpha^{(y)}$, but the sum rule $\sum_{i}  \alpha^{(i)}  = \alpha_\parallel + 2 \alpha_\perp $ still holds. Apparently the molecular quantum state undergoes a drastic change at low $B$ fields when sweeping the magnetic field. Since $\alpha_\parallel$ and  $\alpha_\perp$ are known, we can directly extract the degrees of alignment from the measured polarizabilities $\alpha^{(i)}$ according to Eq.\:(2). From a simultaneous fit, assuming $\langle A_x^2 \rangle = \langle A_y^2 \rangle$ and using the normalization $\langle A_x^2 \rangle + \langle A_y^2 \rangle + \langle A_z^2 \rangle =1$, we obtain $\langle A_{x,y}^2 \rangle = 0.42$ and $\langle A_{z}^2 \rangle = 0.16$. Coupled channel calculations show that in the direct vicinity of the initial state $|2,0\rangle$ ($f=2$, $I=3$, $S=1$) only rotational levels with $R=2$ are present \cite{TiemannPC}. The measured values closely match the calculated degrees of alignment $\langle A_{x,y}^2 \rangle = 0.4286$ and $\langle A_{z}^2 \rangle = 0.1429$ of the states $|2, \pm 2\rangle$, whereas no agreement is found for $|2, \pm 1\rangle$ ($\langle A_{x,y}^2 \rangle = 0.2857$ and $\langle A_{z}^2 \rangle = 0.4286$). How the change in the molecular quantum state comes about is currently still an open question and necessitates further investigation.

In conclusion, we have studied the axis alignment of trapped, ultracold, nonpolar molecules in two predetermined rotational quantum states. For this purpose, we introduced a novel method that relies on the measurement of the polarizability of the molecules along the three spatial axes. Furthermore, we have demonstrated how to apply this method to spectroscopically investigate unknown molecular quantum states.

We have verified that sizeable alignment or antialignment (i.e., $\langle A_{i}^2 \rangle < 1/3$) of the molecular axis can be achieved with spin polarized samples even without any alignment forces. We determined the dynamic polarizabilities $\alpha_\parallel=(8.9 \pm 0.9)\times10^3\:\textrm{a.u.}$ and $\alpha_\perp=(0.9 \pm 0.4)\times10^3\:\textrm{a.u.}$ for the Rb$_2$ vibrational ground state of the  $a^3\Sigma_u^+$ potential at a wavelength of $1064.5\:\textrm{nm}$. The fact that $\alpha_\parallel$ and $\alpha_\perp$ are so different implies that the lattice depth strongly depends on the rotational level. This can be used for filtering, i.e., state selection of molecules. The ability to prepare well-defined rotational states and the successful probing of their alignment paves the way for collisional studies of stereochemical processes. Indeed, we are currently investigating collisions between aligned molecules in a quasi-1D geometry.

\begin{acknowledgments}
The authors would like to thank Eberhard Tiemann,  Olivier Dulieu, and Robert Moszy\'{n}ski for helpful discussions and
information. This work was supported by the German Research Foundation (DFG).  B.D.\ acknowledges support from the Carl-Zeiss-Stiftung.
\end{acknowledgments}

\bibliographystyle{apsprl}

\begin{thebibliography}{22}%
\makeatletter
\providecommand \@ifxundefined [1]{%
 \@ifx{#1\undefined}
}%
\providecommand \@ifnum [1]{%
 \ifnum #1\expandafter \@firstoftwo
 \else \expandafter \@secondoftwo
 \fi
}%
\providecommand \@ifx [1]{%
 \ifx #1\expandafter \@firstoftwo
 \else \expandafter \@secondoftwo
 \fi
}%
\providecommand \natexlab [1]{#1}%
\providecommand \enquote  [1]{``#1''}%
\providecommand \bibnamefont  [1]{#1}%
\providecommand \bibfnamefont [1]{#1}%
\providecommand \citenamefont [1]{#1}%
\providecommand \href@noop [0]{\@secondoftwo}%
\providecommand \href [0]{\begingroup \@sanitize@url \@href}%
\providecommand \@href[1]{\@@startlink{#1}\@@href}%
\providecommand \@@href[1]{\endgroup#1\@@endlink}%
\providecommand \@sanitize@url [0]{\catcode `\\12\catcode `\$12\catcode
  `\&12\catcode `\#12\catcode `\^12\catcode `\_12\catcode `\%12\relax}%
\providecommand \@@startlink[1]{}%
\providecommand \@@endlink[0]{}%
\providecommand \url  [0]{\begingroup\@sanitize@url \@url }%
\providecommand \@url [1]{\endgroup\@href {#1}{\urlprefix }}%
\providecommand \urlprefix  [0]{URL }%
\providecommand \Eprint [0]{\href }%
\@ifxundefined \urlstyle {%
  \providecommand \doi  [0]{\begingroup \@sanitize@url \@doi}%
  \providecommand \@doi [1]{\endgroup \@@startlink {\doibase
  #1}doi:\discretionary {}{}{}#1\@@endlink }%
}{%
  \providecommand \doi  [0]{doi:\discretionary{}{}{}\begingroup
  \urlstyle{rm}\Url }%
}%
\providecommand \doibase [0]{http://dx.doi.org/}%
\providecommand \Doi [0]{\begingroup \@sanitize@url \@Doi }%
\providecommand \@Doi  [1]{\endgroup\@@startlink{\doibase#1}\@@Doi}%
\providecommand \@@Doi [1]{#1\@@endlink}%
\providecommand \selectlanguage [0]{\@gobble}%
\providecommand \bibinfo  [0]{\@secondoftwo}%
\providecommand \bibfield  [0]{\@secondoftwo}%
\providecommand \translation [1]{[#1]}%
\providecommand \BibitemOpen [0]{}%
\providecommand \bibitemStop [0]{}%
\providecommand \bibitemNoStop [0]{.\EOS\space}%
\providecommand \EOS [0]{\spacefactor3000\relax}%
\providecommand \BibitemShut  [1]{\csname bibitem#1\endcsname}%

\bibitem [{\citenamefont {Stapelfeldt}\ \emph {et~al.}(2003)\citenamefont {Stapelfeldt}\ and\ \citenamefont {Seideman}
}]{Stap03}%
  \BibitemOpen
    \bibfield  {author} {\bibinfo {author} {\bibfnamefont {H.}\ \bibnamefont
    {Stapelfeldt}} and
    \bibinfo {author} {\bibfnamefont {T.}\ \bibnamefont
  {Seideman}},\ }
  \Doi
  {10.1103/RevModPhys.75.543} {\bibfield  {journal} {\bibinfo  {journal}
    {Rev. Mod. Phys.} }\textbf {\bibinfo {volume} {75}},\ \bibinfo {pages}
    {543} (\bibinfo {year} {2003})}.
\bibitem [{\citenamefont {Zare}\ \emph {et~al.}(1998)\citenamefont {Zare}
}]{Zare98}%
  \BibitemOpen
    \bibfield  {author} {\bibinfo {author} {\bibfnamefont {R.~N.}\ \bibnamefont
    {Zare}},\ }
  \Doi
  {10.1126/science.279.5358.1875} {\bibfield  {journal} {\bibinfo  {journal}
    {Science} }\textbf {\bibinfo {volume} {279}},\ \bibinfo {pages}
    {1875} (\bibinfo {year} {1998})}.
\bibitem [{\citenamefont {Gijsbertsen}\ \emph {et~al.}(2007)\citenamefont {Gijsbertsen}, \citenamefont {Siu}, \citenamefont {Kling}, \citenamefont {Johnsson}, \citenamefont {Jansen}, \citenamefont {Stolte},\ and\ \citenamefont {Vrakking}
}]{Gijsbertsen07}%
  \BibitemOpen
    \bibfield  {author} {\bibinfo {author} {\bibfnamefont {A.}\ \bibnamefont
    {Gijsbertsen}}, \bibinfo {author} {\bibfnamefont {W.}\ \bibnamefont {Siu}}, \bibinfo {author} {\bibfnamefont {M.~F.}\ \bibnamefont {Kling}}, \bibinfo {author} {\bibfnamefont {P.}\ \bibnamefont {Johnsson}}, \bibinfo {author} {\bibfnamefont {P.}\ \bibnamefont {Jansen}}, \bibinfo {author} {\bibfnamefont {S.}\ \bibnamefont {Stolte}}, and
    \bibinfo {author} {\bibfnamefont {M.~J.~J.}\ \bibnamefont
  {Vrakking}},\ }
  \Doi
  {10.1103/PhysRevLett.99.213003} {\bibfield  {journal} {\bibinfo  {journal}
    {Phys. Rev. Lett.} }\textbf {\bibinfo {volume} {99}},\ \bibinfo {pages}
    {213003} (\bibinfo {year} {2007})}.
\bibitem [{\citenamefont {Ning}\ \emph {et~al.}(2012)\citenamefont {Ning}\ and\ \citenamefont {Seideman}
}]{Ning12}%
  \BibitemOpen
    \bibfield  {author} {\bibinfo {author} {\bibfnamefont {Z.}\ \bibnamefont
    {Ning}} and
    \bibinfo {author} {\bibfnamefont {J.~C.}\ \bibnamefont
  {Polanyi}},\ }
  \Doi
  {10.1063/1.4746803} {\bibfield  {journal} {\bibinfo  {journal}
    {J. Chem. Phys.} }\textbf {\bibinfo {volume} {137}},\ \bibinfo {pages}
    {091706} (\bibinfo {year} {2012})}.
\bibitem [{\citenamefont {Krausz}\ \emph {et~al.}(2009)\citenamefont {Krausz}\ and\ \citenamefont {Ivanov}
}]{Krausz09}%
  \BibitemOpen
    \bibfield  {author} {\bibinfo {author} {\bibfnamefont {F.}\ \bibnamefont
    {Krausz}} and
    \bibinfo {author} {\bibfnamefont {M.}\ \bibnamefont
  {Ivanov}},\ }
  \Doi
  {10.1103/RevModPhys.81.163} {\bibfield  {journal} {\bibinfo  {journal}
    {Rev. Mod. Phys.} }\textbf {\bibinfo {volume} {81}},\ \bibinfo {pages}
    {163} (\bibinfo {year} {2009})}.
\bibitem [{\citenamefont {Baugh}\ \emph {et~al.}(1994)\citenamefont {Baugh}, \citenamefont {Young Kim}, \citenamefont {Cho}, \citenamefont {Pipes}, \citenamefont {Petteway},\ and\ \citenamefont {Fuglesang}
}]{Baugh1994}%
  \BibitemOpen
    \bibfield  {author}
{\bibinfo {author} {\bibfnamefont {D.~A.}\ \bibnamefont {Baugh}},
\bibinfo {author} {\bibfnamefont {D.~Y.}\ \bibnamefont {Kim}},
\bibinfo {author} {\bibfnamefont {V.~A.}\ \bibnamefont {Cho}},
\bibinfo {author} {\bibfnamefont {L.~C.}\ \bibnamefont {Pipes}},
\bibinfo {author} {\bibfnamefont {J.~C.}\ \bibnamefont {Petteway}},
and
    \bibinfo {author} {\bibfnamefont {C.~D.}\ \bibnamefont
  {Fuglesang}},\ }
  \Doi
  {}
{\bibfield  {journal} {\bibinfo  {journal}
    {Chem. Phys. Lett.} }\textbf {\bibinfo {volume} {219}},\ \bibinfo {pages}
    {207} (\bibinfo {year} {1994})}.
\bibitem [{\citenamefont {Mukherjee}\ \emph {et~al.}(2011)\citenamefont {Mukherjee}\ and\ \citenamefont {Zare}
}]{Mukherjee11}%
  \BibitemOpen
    \bibfield  {author} {\bibinfo {author} {\bibfnamefont {N.}\ \bibnamefont
    {Mukherjee}} and
    \bibinfo {author} {\bibfnamefont {R.~N.}\ \bibnamefont
  {Zare}},\ }
  \Doi
  {10.1063/1.3599711} {\bibfield  {journal} {\bibinfo  {journal}
    {J. Chem. Phys.} }\textbf {\bibinfo {volume} {135}},\ \bibinfo {pages}
    {024201} (\bibinfo {year} {2011})}.
\bibitem [{\citenamefont {Friedrich}\ \emph {et~al.}(1992)\citenamefont {Friedrich}\ and\ \citenamefont {Herschbach}
}]{Friedrich92}%
  \BibitemOpen
    \bibfield  {author} {\bibinfo {author} {\bibfnamefont {B.}\ \bibnamefont
    {Friedrich}} and
    \bibinfo {author} {\bibfnamefont {D.~R.}\ \bibnamefont
  {Herschbach}},\ }
  \Doi
  {10.1007/BF01436600} {\bibfield  {journal} {\bibinfo  {journal}
    {Z. Phys. D} }\textbf {\bibinfo {volume} {24}},\ \bibinfo {pages}
    {25} (\bibinfo {year} {1992})}.
\bibitem [{\citenamefont {Lemeshko}\ \emph {et~al.}(2013)\citenamefont {Lemeshko}, \citenamefont {Krems}, \citenamefont {Doyle},\ and\ \citenamefont {Kais}
}]{Lemeshko13}%
  \BibitemOpen
    \bibfield  {author} {\bibinfo {author} {\bibfnamefont {M.}\ \bibnamefont
    {Lemeshko}}, \bibinfo {author} {\bibfnamefont {R.~V.}\ \bibnamefont {Krems}}, \bibinfo {author} {\bibfnamefont {J.~M.}\ \bibnamefont {Doyle}}, and
    \bibinfo {author} {\bibfnamefont {S.}\ \bibnamefont
  {Kais}},\ }
  \Doi
  {10.1080/00268976.2013.813595} {\bibfield  {journal} {\bibinfo  {journal}
    {Molecular Physics} }\textbf {\bibinfo {volume} {111}},\ \bibinfo {pages}
    {1648} (\bibinfo {year} {2013})}.
\bibitem [{\citenamefont {Krems}\ \emph {et~al.}(2008)\citenamefont {Krems}
}]{Krems08}%
  \BibitemOpen
    \bibfield  {author} {\bibinfo {author} {\bibfnamefont {R.~V.}\ \bibnamefont
    {Krems}},\ }
  \Doi
  {10.1039/B802322K} {\bibfield  {journal} {\bibinfo  {journal}
    {Phys. Chem. Chem. Phys.} }\textbf {\bibinfo {volume} {10}},\ \bibinfo {pages}
    {4079} (\bibinfo {year} {2008})}.
\bibitem [{\citenamefont {de Miranda}\ \emph {et~al.}(2011)\citenamefont {de Miranda}, \citenamefont {Chotia}, \citenamefont {Neyenhuis}, \citenamefont {Wang}, \citenamefont {Qu\'{e}m\'{e}ner}, \citenamefont {Ospelkaus}, \citenamefont {Bohn}, \citenamefont {Ye},\ and\ \citenamefont {Jin}
}]{Miranda11}%
  \BibitemOpen
    \bibfield  {author} {\bibinfo {author} {\bibfnamefont {M.~H.~G.}\ \bibnamefont
    {de Miranda}}, \bibinfo {author} {\bibfnamefont {A.}\ \bibnamefont {Chotia}}, \bibinfo {author} {\bibfnamefont {B.}\ \bibnamefont {Neyenhuis}}, \bibinfo {author} {\bibfnamefont {D.}\ \bibnamefont {Wang}}, \bibinfo {author} {\bibfnamefont {G.}\ \bibnamefont {Qu\'{e}m\'{e}ner}}, \bibinfo {author} {\bibfnamefont {S.}\ \bibnamefont {Ospelkaus}}, \bibinfo {author} {\bibfnamefont {J.~L.}\ \bibnamefont {Bohn}}, \bibinfo {author} {\bibfnamefont {J.}\ \bibnamefont {Ye}}, and
    \bibinfo {author} {\bibfnamefont {D.~S.}\ \bibnamefont
  {Jin}},\ }
  \Doi
  {10.1038/nphys1939} {\bibfield  {journal} {\bibinfo  {journal}
    {Nat. Phys.} }\textbf {\bibinfo {volume} {7}},\ \bibinfo {pages}
    {502} (\bibinfo {year} {2011})}.
\bibitem [{\citenamefont {Neyenhuis}\ \emph {et~al.}(2012)\citenamefont {Neyenhuis}, \citenamefont {Yan}, \citenamefont {Moses}, \citenamefont {Covey}, \citenamefont {Chotia}, \citenamefont {Petrov}, \citenamefont {Kotochigova}, \citenamefont {Ye},\ and\ \citenamefont {Jin}
}]{Neyenhuis2012}%
  \BibitemOpen
    \bibfield  {author} {\bibinfo {author} {\bibfnamefont {B.}\ \bibnamefont
    {Neyenhuis}}, \bibinfo {author} {\bibfnamefont {B.}\ \bibnamefont {Yan}}, \bibinfo {author} {\bibfnamefont {S.~A.}\ \bibnamefont {Moses}}, \bibinfo {author} {\bibfnamefont {J.~P.}\ \bibnamefont {Covey}}, \bibinfo {author} {\bibfnamefont {A.}\ \bibnamefont {Chotia}}, {\bibfnamefont {A.}\ \bibnamefont {Petrov}}, {\bibfnamefont {S.}\ \bibnamefont {Kotochigova}}, {\bibfnamefont {J.}\ \bibnamefont {Ye}}, and
    \bibinfo {author} {\bibfnamefont {D.~S.}\ \bibnamefont
  {Jin}},\ }
  \Doi
  {10.1103/PhysRevLett.109.230403} {\bibfield  {journal} {\bibinfo  {journal}
    {Phys. Rev. Lett.} }\textbf {\bibinfo {volume} {109}},\ \bibinfo {pages}
    {230403} (\bibinfo {year} {2012})}.
\bibitem [{\citenamefont {Rutkowski}\ \emph {et~al.}(2004)\citenamefont {Rutkowski}\ and\ \citenamefont {Zacharias}
}]{Rutkowski04}%
  \BibitemOpen
    \bibfield  {author} {\bibinfo {author} {\bibfnamefont {M.}\ \bibnamefont
    {Rutkowski}} and
    \bibinfo {author} {\bibfnamefont {H.}\ \bibnamefont
  {Zacharias}},\ }
  \Doi
  {10.1016/j.chemphys.2004.01.009} {\bibfield  {journal} {\bibinfo  {journal}
    {Chem. Phys.} }\textbf {\bibinfo {volume} {301}},\ \bibinfo {pages}
    {189} (\bibinfo {year} {2004})} and
  \Doi
    {10.1016/j.chemphys.2004.10.038} {\bibfield  {journal} {\bibinfo  {journal}
      {Chem. Phys.} }\textbf {\bibinfo {volume} {310}},\ \bibinfo {pages}
      {321} (\bibinfo {year} {2005})}.
\bibitem [{\citenamefont {Bartlett}\ \emph {et~al.}(2010)\citenamefont {Bartlett}, \citenamefont {Jankunas}, \citenamefont {Zare}, \citenamefont {Sofikitis}, \citenamefont {Rakitzis},\ and\ \citenamefont {Harrison}
}]{Bartlett10}%
  \BibitemOpen
    \bibfield  {author} {\bibinfo {author} {\bibfnamefont {N.~C.-M.}\ \bibnamefont
    {Bartlett}}, \bibinfo {author} {\bibfnamefont {J.}\ \bibnamefont {Jankunas}}, \bibinfo {author} {\bibfnamefont {R.~N.}\ \bibnamefont {Zare}}, and
    \bibinfo {author} {\bibfnamefont {J.~A.}\ \bibnamefont
  {Harrison}},\ }
  \Doi
  {10.1039/C0CP00518E} {\bibfield  {journal} {\bibinfo  {journal}
    {Phys. Chem. Chem. Phys.} }\textbf {\bibinfo {volume} {12}},\ \bibinfo {pages}
    {15689} (\bibinfo {year} {2010})}.
\bibitem [{tob()}]{coupling}%
 \BibitemOpen
 \href@noop {} {}\bibinfo {note} {
 Interestingly,  $J$ ($\vec{J} =  \vec{R} + \vec{S}$) is in general not a good quantum number 
 for Rb$_2$ in state $a^3\Sigma_u^+$ because the electron spin $\vec{S}$ is only weakly coupled to $\vec{R}$ via second-order spin-orbit interaction but strongly to the nuclear spin $\vec{I}$ via hyperfine interaction \cite{Strauss2010}. Thus $f$ ($\vec{f} =  \vec{S} + \vec{I}$) is good.
The weak spin-orbit interaction tends to couple $\vec{f}$ and $\vec{R}$ to form the total angular momentum $\vec{F}$, but already a small Zeeman interaction due to a $B$ field of a few tens of gauss leads again to decoupling, especially for deeply bound molecules.}
\bibitem [{\citenamefont {Lang}\ \emph {et~al.}(2008)\citenamefont {Lang}, \citenamefont {Winkler}, \citenamefont {Strauss}, \citenamefont {Grimm},\ and\ \citenamefont {Hecker Denschlag}
}]{Lang2008}%
  \BibitemOpen
    \bibfield  {author} {\bibinfo {author} {\bibfnamefont {F.}\ \bibnamefont
    {Lang}}, \bibinfo {author} {\bibfnamefont {K.}\ \bibnamefont {Winkler}}, \bibinfo {author} {\bibfnamefont {C.}\ \bibnamefont {Strauss}}, \bibinfo {author} {\bibfnamefont {R.}\ \bibnamefont {Grimm}}, and
    \bibinfo {author} {\bibfnamefont {J.}\ \bibnamefont
  {Hecker Denschlag}},\ }
  \Doi
  {10.1103/PhysRevLett.101.133005} {\bibfield  {journal} {\bibinfo  {journal}
    {Phys. Rev. Lett.} }\textbf {\bibinfo {volume} {101}},\ \bibinfo {pages}
    {133005} (\bibinfo {year} {2008})}.
\bibitem [{\citenamefont {Strauss}\ \emph {et~al.}(2010)\citenamefont {Strauss}, \citenamefont {Takekoshi}, \citenamefont {Lang}, \citenamefont {Winkler}, \citenamefont {Grimm},\ and\ \citenamefont {Hecker Denschlag}, \citenamefont {Tiemann}
}]{Strauss2010}%
  \BibitemOpen
    \bibfield  {author} {\bibinfo {author} {\bibfnamefont {C.}\ \bibnamefont
    {Strauss}}, \bibinfo {author} {\bibfnamefont {T.}\ \bibnamefont {Takekoshi}}, \bibinfo {author} {\bibfnamefont {F.}\ \bibnamefont {Lang}}, \bibinfo {author} {\bibfnamefont {K.}\ \bibnamefont {Winkler}}, \bibinfo {author} {\bibfnamefont {R.}\ \bibnamefont {Grimm}}, and
    \bibinfo {author} {\bibfnamefont {J.}\ \bibnamefont
  {Hecker Denschlag}}, \bibinfo {author} {\bibfnamefont {E.}\ \bibnamefont {Tiemann}},\ }
  \Doi
  {10.1103/PhysRevA.82.052514} {\bibfield  {journal} {\bibinfo  {journal}
    {Phys. Rev. A} }\textbf {\bibinfo {volume} {82}},\ \bibinfo {pages}
    {052514} (\bibinfo {year} {2010})}.
\bibitem [{\citenamefont {Hecker Denschlag}\ \emph {et~al.}(2002)\citenamefont {Hecker Denschlag}, \citenamefont {H\"{a}ffner}, \citenamefont {McKenzie}, \citenamefont {Browaeys}, \citenamefont {Cho},
    \citenamefont {Helmerson}, \citenamefont {Rolston},\ and\ \citenamefont {Phillips}
}]{Denschlag02}%
  \BibitemOpen
    \bibfield  {author} {\bibinfo {author} {\bibfnamefont {J.}\ \bibnamefont
    {Hecker Denschlag}},
    \bibinfo {author} {\bibfnamefont {H.}\ \bibnamefont {H\"{a}ffner}},
    \bibinfo {author} {\bibfnamefont {C.}\ \bibnamefont {McKenzie}},
    \bibinfo {author} {\bibfnamefont {A.}\ \bibnamefont {Browaeys}},
    \bibinfo {author} {\bibfnamefont {D.}\ \bibnamefont {Cho}},
    \bibinfo {author} {\bibfnamefont {C.}\ \bibnamefont {Helmerson}},
    \bibinfo {author} {\bibfnamefont {S.}\ \bibnamefont {Rolston}},
     and
    \bibinfo {author} {\bibfnamefont {W.~D.}\ \bibnamefont
  {Phillips}},\ }
  \Doi
  { doi:10.1088/0953-4075/35/14/307} {\bibfield  {journal} {\bibinfo  {journal}
    {J. Phys. B: At. Mol. Opt. Phys.} }\textbf {\bibinfo {volume} {35}},\ \bibinfo {pages}
    {3095} (\bibinfo {year} {2002})}.
\bibitem [{\citenamefont {Danzl}\ \emph {et~al.}(2010)\citenamefont {Danzl}, \citenamefont {Mark}, \citenamefont {Haller}, \citenamefont {Gustavsson}, \citenamefont {Hart}, \citenamefont {Aldegunde}, \citenamefont {Hutson},\ and\ \citenamefont {N\"{a}gerl}
}]{Danzl2010}%
  \BibitemOpen
    \bibfield  {author} {\bibinfo {author} {\bibfnamefont {J.~G.}\ \bibnamefont
    {Danzl}}, \bibinfo {author} {\bibfnamefont {M.~J.}\ \bibnamefont {Mark}}, \bibinfo {author} {\bibfnamefont {E.}\ \bibnamefont {Haller}}, \bibinfo {author} {\bibfnamefont {M.}\ \bibnamefont {Gustavsson}}, \bibinfo {author} {\bibfnamefont {R.}\ \bibnamefont {Hart}}, {\bibfnamefont {J.}\ \bibnamefont {Aldegunde}}, {\bibfnamefont {J.~M.}\ \bibnamefont {Hutson}}, and
    \bibinfo {author} {\bibfnamefont {H.-C.}\ \bibnamefont
  {N\"{a}gerl}},\ }
  \Doi
  {10.1038/nphys1533} {\bibfield  {journal} {\bibinfo  {journal}
    {Nat. Phys.} }\textbf {\bibinfo {volume} {6}},\ \bibinfo {pages}
    {265} (\bibinfo {year} {2010})}.
\bibitem [{\citenamefont {Safronova}\ \emph {et~al.}(2004)\citenamefont {Safronova}, \citenamefont {Williams},\ and\ \citenamefont {Clark}
}]{Safronova2004}%
  \BibitemOpen
    \bibfield  {author}
{\bibinfo {author} {\bibfnamefont {M.~S.}\ \bibnamefont {Safronova}},
\bibinfo {author} {\bibfnamefont {C.~J.}\ \bibnamefont {Williams}},
 and
    \bibinfo {author} {\bibfnamefont {C.~W.}\ \bibnamefont
  {Clark}},\ }
  \Doi
  {10.1103/PhysRevA.69.022509} {\bibfield  {journal} {\bibinfo  {journal}
    {Phys. Rev. A} }\textbf {\bibinfo {volume} {69}},\ \bibinfo {pages}
    {022509} (\bibinfo {year} {2004})}.
\bibitem [{tob()}]{TomzaPC}%
 \BibitemOpen
 \href@noop {} {}\bibinfo {note} {M. Tomza and R. Moszy\'{n}ski (private communication).}%
\bibitem [{\citenamefont {Tomza}\ \emph {et~al.}(2002)\citenamefont {Tomza}, \citenamefont {Skomorowski}, \citenamefont {Musia\l}, \citenamefont {Gonz\'{a}lez-F\'{e}rez}, \citenamefont {Koch},\ and\ \citenamefont {Moszynski}
}]{Tomza2013}%
  \BibitemOpen
    \bibfield  {author} {\bibinfo {author} {\bibfnamefont {M.}\ \bibnamefont
    {Tomza}},
    \bibinfo {author} {\bibfnamefont {W.}\ \bibnamefont {Skomorowski}},
    \bibinfo {author} {\bibfnamefont {M.}\ \bibnamefont {Musia\l}},
    \bibinfo {author} {\bibfnamefont {R.}\ \bibnamefont {Gonz\'{a}lez-F\'{e}rez}},
    \bibinfo {author} {\bibfnamefont {C.~P.}\ \bibnamefont {Koch}},
     and
    \bibinfo {author} {\bibfnamefont {R.}\ \bibnamefont
  {Moszynski}},\ }
  \Doi
  { doi::10.1080/00268976.2013.793835} {\bibfield  {journal} {\bibinfo  {journal}
    {Molecular Physics} }\textbf {\bibinfo {volume} {111}},\ \bibinfo {pages}
    {1781} (\bibinfo {year} {2013})}.
\bibitem [{tob()}]{supplemental}%
 \BibitemOpen
 \href@noop {} {}\bibinfo {note} {See Supplemental Material at http://link.aps.org/
 supplemental/10.1103/PhysRevLett.113.233004, which includes potential energy curves from Ref. \cite{Lozeille2006}, for a qualitative explanation of the large difference of the molecular polarizabilities $\alpha_\parallel$ and $\alpha_\perp$.}%
\bibitem [{\citenamefont {Lozeille}\ \emph {et~al.}(2006)\citenamefont {Lozeille}, \citenamefont {Williams}, \citenamefont {Fioretti}, \citenamefont {Gabbanini}, \citenamefont {Huang}, \citenamefont {Pechkis}, \citenamefont {Wang}, \citenamefont {Gould}, \citenamefont {Eyler}, \citenamefont {Stwalley}, \citenamefont {Aymar},\ and\ \citenamefont {Dulieu}
}]{Lozeille2006}%
  \BibitemOpen
    \bibfield  {author}
{\bibinfo {author} {\bibfnamefont {J.}\ \bibnamefont {Lozeille}},
\bibinfo {author} {\bibfnamefont {A.}\ \bibnamefont {Fioretti}},
\bibinfo {author} {\bibfnamefont {C.}\ \bibnamefont {Gabbanini}},
\bibinfo {author} {\bibfnamefont {Y.}\ \bibnamefont {Huang}},
\bibinfo {author} {\bibfnamefont {H.~K.}\ \bibnamefont {Pechkis}},
\bibinfo {author} {\bibfnamefont {D.}\ \bibnamefont {Wang}},
\bibinfo {author} {\bibfnamefont {P.~L.}\ \bibnamefont {Gould}},
\bibinfo {author} {\bibfnamefont {E.~E.}\ \bibnamefont {Eyler}},
\bibinfo {author} {\bibfnamefont {W.~C.}\ \bibnamefont {Stwalley}},
\bibinfo {author} {\bibfnamefont {M.}\ \bibnamefont {Aymar}},
and
    \bibinfo {author} {\bibfnamefont {O.}\ \bibnamefont
  {Dulieu}},\ }
  \Doi
  {10.1103/PhysRevA.69.022509}
{\bibfield  {journal} {\bibinfo  {journal}
    {Eur. Phys. J. D} }\textbf {\bibinfo {volume} {39}},\ \bibinfo {pages}
    {261} (\bibinfo {year} {2006})}.
\bibitem [{tob()}]{TiemannPC}%
 \BibitemOpen
 \href@noop {} {}\bibinfo {note} {E. Tiemann (private communication).}%

\end{thebibliography}

\end{document}


\title{Supplemental material - Probing the axis alignment of an ultracold spin-polarized $\textrm{Rb}_2$ molecule}
\author{Markus Dei{\ss}}
\author{Bj\"orn Drews}
\author{Benjamin Deissler}
\author{Johannes Hecker Denschlag}
\affiliation{Institut f\"ur Quantenmaterie and Center for Integrated Quantum Science and Technology IQ$^{ST}$, Universit\"at Ulm, 89069 Ulm, Germany}

\date{\today}

\pacs{33.15.Kr, 33.15.Bh, 37.10.Pq, 67.85.-d}

\maketitle

\begin{figure}[h]
\renewcommand{\thefigure}{S1}
\includegraphics[width=8.0cm] {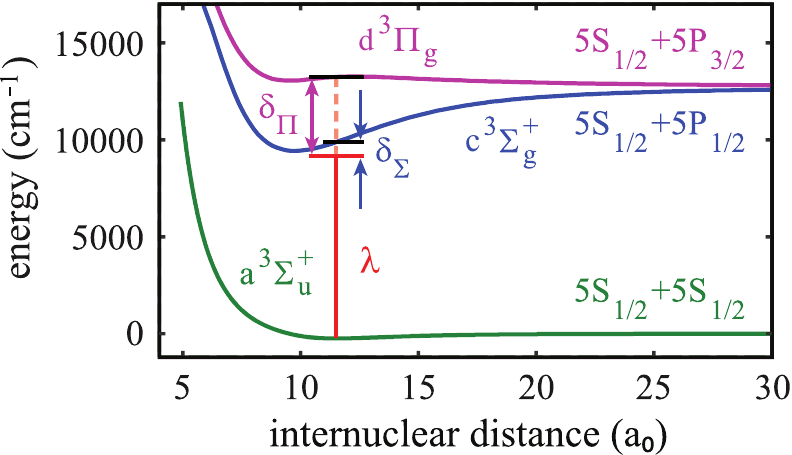}
\caption{(Color online). Detunings $\delta_\Sigma$ and $\delta_\Pi$ of the lattice laser at $\lambda=1064.5\:\textrm{nm}$ from the two most relevant transitions, $a^3\Sigma_u^+-c^3\Sigma_g^+$ and $a^3\Sigma_u^+-d^3\Pi_g$, that determine the polarizabilities $\alpha_\parallel$  and $\alpha_\perp$, respectively. The energy reference is given by the asymptote of the $a^3\Sigma_u^+$ potential. The potential curves are taken from \cite{Lozeille2006}.
\label{figS1}
}
\end{figure}

Owing to selection rules for $\pi$- and $\sigma$-transitions, $\alpha_\parallel$ ($\alpha_\perp$) is linked to $\Sigma-\Sigma$  ($\Sigma-\Pi$) transitions. As can be seen from Fig.\:\ref{figS1}, the detuning of our lattice laser from the Franck-Condon point on the $d^3\Pi_g$ potential curve is about 6 times larger than from that of the $c^3\Sigma_g^+$ potential curve, explaining the large difference of $\alpha_\parallel$ and $\alpha_\perp$ to first order.

\bibliographystyle{apsprl}